\documentclass[preprint,showpacs,preprintnumbers,amsmath,amssymb]{revtex4}
\usepackage{graphicx}
\usepackage{dcolumn}
\usepackage{bm}
\usepackage[usenames]{color}
\newcommand{\be}{\begin{equation}}
\newcommand{\en}{\end{equation}}
\newcommand{\bea}{\begin{eqnarray}}
\newcommand{\ena}{\end{eqnarray}}

\usepackage{perpage} 
\MakePerPage{footnote} 

\begin{document}


\title{ Reconstruction and exact solutions for cosmological
perturbations from a generalized gravity theory.}
\author{Jos\'e Antonio Belinch\'on}

\email{jose.belinchon@uda.cl} \affiliation{ Departamento de
Matem\'aticas, Facultad de Ingenier\'ia, Universidad de Atacama,
Avenida Copayapu 485, Copiap\'{o}, Chile.}

\author{Carlos Gonz\'{a}lez}
\email{carlos.gonzalez@uda.cl} \affiliation{ Departamento de F\'{\i}sica, Universidad de Atacama, Avenida Copayapu 485,
Copiap\'{o}, Chile.}

\author{Ram\'on Herrera}

\email{ramon.herrera@pucv.cl} \affiliation{ Instituto de
F\'{\i}sica, Pontificia Universidad Cat\'{o}lica de
Valpara\'{\i}so, Avenida Brasil 2950, Casilla 4059,
Valpara\'{\i}so, Chile.}

\date{\today}
\begin{abstract}
Scalar and tensor cosmological perturbations during  an
inflationary universe scenario in the context of the  a
generalized gravity theory are studied. This analyze is carried
out
 considering an ansatz on  the variables associated to scalar and
tensor perturbation ($z_s$ and $z_t$) in the Jordan frame.  In
this context,
 we analyze two different Ansatze for the ratio $z_s/z_t$, and we
 study
 in great detail the  analytical and exact
solutions for the cosmological perturbations together with the
corresponding  reconstruction of the background variables. Recent
observational data from the Planck 2018 results are employed to
constrain the parameters of each of the models.

\end{abstract}

\maketitle

\section{Introduction}

It is well known that the inflationary stage is to date  an
excellent candidate for solving the long standing problems of the
 Big Bang model (horizon, flatness, monopoles, etc.) during the early universe\cite{prob,
 prob2}. Nevertheless, the biggest characteristic of the inflationary
 universe is that it furnishes  a causal interpretation
to explain  the observed  anisotropy of the cosmic microwave
background (CMB) radiation\cite{cmb,Planck2015} and also inflation
gives account of the distribution of large scale structures
\cite{ls}.

In order to describe the inflationary epoch, the scalar field
plays a fundamental role in the dynamic of the early universe. In
this sense, the scalar field or inflaton can interact essentially
with other fields and also with the gravitational sector. In the
context of  the different inflationary  models that give account
of the  evolution of the early universe, we can distinguish those
the models that use a scalar tensor theory in order to produce the
accelerated expansion of the universe. The scalar to tensor theory
of gravitation can be considered as an extension to the General
Relativity (GR), where a  scalar field  couples with the
gravitational sector produces  a dynamic evolution on the
gravitational constant or Planck mass. In this context, the scalar-tensor
theory of gravity  as alternative theory is the most well studied that exist in
the literature.
Originally,  this theory
was proposed in the works of Kaluza and Klein, Jordan\cite{Jor},
Brans and Dicke\cite{BD} and its generalization by Bergamann and
Wagoner\cite{BW}. In particular,   the Jordan Brans Dicke (JBD)
theory is formulated by the presence of a  scalar field (massless)
couples with the gravitational sector from the scalar curvature
applied to the cosmology. In the context of  radiation problems
assuming the non-minimal coupling with the Ricci scalar was
analyzed in Ref.\cite{RP}. It is well known that the scalar tensor theories
emerge clearly from effective theories of higher dimensional theories under a
dimensional reduction, e.g. Kaluza-Klein, string\cite{Clifton:2011jh}. In order to describe the inflationary epoch,
different inflationary models have been
studied in the framework of the non-minimal coupling to gravity,
called the Jordan frame, see e.g.\cite{1,2,3}. In relation to the
 Jordan and Einstein  frames in Ref.\cite{nr} was analyzed a frame independent
 classification from the point of view of the observable
 parameters;  the scalar spectral index $n_s$ and the
tensor to scalar ratio $r$ , see also \cite{Kaiser:1994vs}. For a
review on the scalar-tensor theory and its context in the
inflationary stage, see Refs.\cite{p1,p2,p3}.

On the other hand, the study of  the  cosmological perturbations,
such as the power spectrum of the scalar  and tensor
perturbations coming   from quantum fluctuations of the vacuum in
the early  universe was developed in Ref.\cite{ml}. During the
early universe, these fluctuations have a small amplitudes
approximately of the Planck scale, and product of the accelerated
expansion of the universe, these fluctuations
 are amplified  to scales farther  galactic scales.

 In this context,   analytical and exact  solutions from  the
   background  and   perturbative equations (scalar-tensor)  in the case of  a standard scalar field can be
   obtained in the Einstein frame.
   A
   de Sitter universe is an example of this, here
    a constant potential associated to scalar field  is
obtained  from the  background\cite{prob},
and an exact results can be found from the perturbative equations
 for the scalar and tensor  power spectrums
 \cite{ml}. Another example corresponds to
 an expansion power-law or power-law inflation. For the case of the power-law
 we can
found an exact results for the background equations  giving as result  an exponential
potential\cite{pl} and we can also got an exact calculation for the cosmological
perturbations, see Ref.\cite{st}.  In this sense, the reconstruction of the exact and analytical  background variables
from the analysis of the perturbative equations of the cosmological
perturbations has been considered by several authors\cite{mukhanov}. In particular  considering  the Einstein framework, an
 exact reconstruction of the background can be
obtained in the specific case in which
 the effective potential\cite{mukhanov} of
perturbative equations is zero (called the Easther model)\cite{Ea}. However, this
condition on the effective potential gives rise to
 a scalar spectral index $n_S\equiv 3$ (blue tilt),  and this value is
 disapproved  by
  the observational data.  By considering  this idea, the authors in ref. \cite{martin}
  consider an anzatz for the
effective potential of the scalar
perturbative equation in terms of the conformal time assuming a standard scalar field. Here, analytical solutions
were found for the scalar and tensor power spectrums,
 however, the reconstruction of  background (inflaton potential) was obtained
numerically.  Similarly, assuming an extended ansatz for the
effective potential of the scalar perturbative equation in the
framework of a tachyon field the authors in
ref.\cite{Herrera:2015udk} found analytical expressions for the
scalar and tensor power spectrums, considering that  the tachyon
field  slowly changes with the time during inflation. Here, the
model is well supported by the Planck data only if the effective
speed of sound $c_s$ associated the tachyonic field is $c_s\simeq
1$
 and the effective potential of the scalar perturbative equation is similar to
 the power-law\cite{Herrera:2015udk}.  In this line,
 some approximations to the cosmological
perturbations in order to obtain analytical expressions were
developed in Refs.\cite{A1, A2} (see also
ref.\cite{Boisseau:2000pr} for the reconstruction during the
present acceleration). Also, in order to solve the scalar
perturbative equation in the Einstein frame, the  authors in
Ref.\cite{Choe:2004zg}
 developed a new formalism, in which the considered methodology  corresponds to  Green's
function method to calculate the power spectrum of the curvature
perturbations produced during inflation, see
also\cite{Wei:2004xx}.

In the following,  we analyze the possibility to reconstruct
 exactly  the background variables and scalar and tensor power
 spectrums in   the frame of a  generalized scalar to tensor theory.  We
 propound this possibility considering a new methodology from an
 ansatz between the scalar and tensor variables ($z_s$ and $z_t$) in terms  of the
conformal time.  In this form, we study how this relation
influences the dynamics of the scalar and tensor perturbations and
the background variables (scale factor, scalar potential and
coupling function) in a  generalized scalar to tensor theory.
Since our intention is to reconstruct exactly the background
variables starting from an already exact solution for cosmological
perturbations, we will see that it is not necessary to consider
slow roll assumption into our treatment. Also, we will explorer
the parameter spaces from  the cosmological perturbations and how
these parameters are constrained from the observational
data\cite{Plancku}.

The outline of the paper goes as follow: In the sections II y III
we give a brief  description of the background equations and the
cosmological perturbations in a generalized  scalar to tensor
theory of gravitation. In the section IV, we study a specify case
in which the scalar and tensor variables are proportional (model
I) and as an example we consider the power law expansion.
 In the section V, we analyze an
explicit expressions for  the scalar and tensor variables as
function of the conformal time (model II),  in order to
reconstruct the background and power  spectrums quantities. Here,
we confine the parameter spaces from Planck 2018 data.
 In section VI exhibits our results and
conclusions. We chose units so that $c=\hbar=8\pi G=1$.

\section{ Background equations}
In this section  we give a brief description of the background
equations, considering a generalized scalar tensor theory. In this
context, we start with the effective action for the scalar-tensor
gravitational model  in the Jordan frame given by\cite{J1}
\begin{equation}
S=\int d^4x \sqrt{-g}\left[ \frac{F(\varphi)}{2} R
+\frac{\omega(\varphi)}{2}g^{\mu\nu}\varphi_{;\mu}\varphi_{;\nu} - V
(\varphi)\right], \label{eq1}
\end{equation}
where the quantity $F(\varphi)$ corresponds to arbitrary function
that describes the coupling between the scalar field $\varphi$ and
the geometry from  the Ricci scalar $R$, $\omega(\varphi)$ denotes
a generic function coupling the scalar field to its kinetic energy
density and $V(\varphi)$ corresponds to the effective potential
associated to the scalar field.

 By considering  the  variation of the effective action given by  Eq.(\ref{eq1}) with respect to the two fields
  $g_{\mu\nu}$ and $\varphi $, we obtain the  background equations.
Let us introduce into the action (\ref{eq1}) a spatially  flat
Friedmann-Lemaitre-Robertson-Walker (FLRW) metric, defined as
 \be
 ds^2=dt^2-a^2(t)\,dx^{i}dx_{i}
 =a^2(\eta)[d\eta^2-dx^{i}dx_{i}],\label{eq2}
 \en
 where the new time $\eta$ denotes  the conformal time defined as $dt=a d\eta$, the quantities  $dx^{i}dx_{i}$
 and $a(t)$ correspond to the flat three-surface and scale factor, respectively.

By assuming a spatially homogeneous  scalar field $\varphi(\eta)$,
then
 the corresponding Friedmann equation reads

\begin{equation}
3\mathcal{H}^{2}=\frac{\omega\varphi^{\prime
2}}{2F}+\frac{a^{2}V}{F}-3\mathcal{H}\frac{F^{\prime}}{F},\label{eq3}
\end{equation}
and the Raychaudhuri equation for this theory can be written as
\begin{equation}
\mathcal{H}^{2}-\mathcal{H}^{\prime}=\frac{\omega\varphi^{\prime
2}}{2F}+\frac{F^{\prime\prime}}{2F}-\frac{\mathcal{H}F^{\prime}}{F}.\label{eq4}
\end{equation}
Here the Hubble rate in the conformal time is defined as
$\mathcal{H}\equiv a^{\prime}/a$ and the scalar Ricci from metric
(\ref{eq2}) is given by
$R=\frac{6}{a^2}(\mathcal{H}^2+\mathcal{H}^{\prime})$. In the
following, the prime $^{\prime}$ denotes the derivative with
respect to conformal time $\eta$.

The dynamical equation of motion for the scalar  field from action
(\ref{eq1}) is given by
\begin{equation}
\frac{1}{2\omega}(a^2 F_{,\varphi}R -\omega_{,\varphi}\varphi^{\prime 2}-2a^2 V_{,\varphi})-2\mathcal{H}\varphi^{\prime}-\varphi^{\prime\prime}=0,\label{eq5}
\end{equation}
where the subscript ($_{,\varphi})$ denotes the derivative with
respect to scalar field $\varphi$.

By combining Eqs.(\ref{eq3}) and (\ref{eq4}) we find that the
effective potential can be rewritten as
\begin{equation}
V=\frac{1}{a^2}\left[F(\mathcal{H}^{\prime}+2\mathcal{H}^{2})+2\mathcal{H}
F^{\prime}+\frac{F^{\prime\prime}}{2}\right].\label{potV}
\end{equation}
Here,  we note that the effective potential can be obtained in
terms of the time (or scalar field) once known  the scale factor
and the arbitrary function $F$.

\section{Perturbations}
In this section we will analyze the scalar and tensor
perturbations for the case of the generalized scalar tensor
theory.  We consider that the   general metric perturbations is
given by \be ds^2=a^{2}(\eta)[(1+\Phi)d\eta ^2-2\partial_i B
dx^id\eta -[(1-2\Psi)\delta_{ij}+2\partial_i\partial_j
E]dx^idx^j]. \label{me} \en From this metric  and in order to
describe  the scalar-type structure we need the four variables
$\Phi$, $B$, $\Psi$ and $E$, which are functions of the time and
space coordinates\cite{Bardeen}.

On the other hand, we also consider that the perturbation
associated to the field $\varphi(t,\vec{x} )$ is given by
$\varphi(t,\vec{x} )=\varphi(t)+\delta\varphi(t,\vec{x} )$, where
$\varphi(t)$ is the homogeneous background field satisfying
(\ref{eq5}) and $\delta\varphi(t,\vec{x} )$ corresponds to the
perturbation in space and time of the scalar field.

Introducing  the intrinsic curvature perturbation $\mathcal{R}$ of
the comoving hypersurfaces, that gives account of the scalar
perturbations defined by
\begin{equation}
\mathcal{R} \equiv \Psi + \mathcal{H} \frac{\delta
\varphi}{\varphi^{\prime}}\,,\label{eq8}
\end{equation}
then considering this intrinsic curvature the first order
perturbed equations can be compacted to \cite{noh}
\begin{equation}
\frac{1}{a^{3}Q_{s}}\frac{d}{dt}{(a^{3}Q_{s}\dot{{\cal{R}}})}
+ \frac{k^{2}}{a^{2}}{\cal{R}}=0,\label{eq9}
\end{equation}
where the function $Q_s$ associated to the scalar perturbation in
the Jordan frame  is defined as \cite{noh}
\begin{equation}
Q_{s}=\frac{ \frac{3F^{\prime 2}}{2F}+\omega\varphi^{\prime 2}
}{(\mathcal{H}+\frac{F^{\prime}}{2F} )^{2}}.\label{eq10}
\end{equation}

Other important quantity to describe the perturbations  is the
gauge-invariant potential $v$, that relates the background
variables with the intrinsic curvature and is defined as
$v=z_s\mathcal{R}$ or in the Fourier modes
$v_{k}=z_{s}{\mathcal{R}_{k}}$. Here $k=|\boldsymbol{k}|$ denotes  the modulus of the wavenumber.
 In this context,  the new variable
$z_s$ associated to scalar perturbations in the framework of the
scalar tensor theory is defined as\cite{noh}
\begin{equation}
z_{s}=a\sqrt{Q_s}. \label{eq11}
\end{equation}

In this form, the  first order perturbed equation (\ref{eq9})
considering  the Fourier modes $v_{k}$ can be rewritten as
\cite{shinji,noh}
\begin{equation}
v_{k}^{\prime\prime} +
\left(k^{2}-\frac{z_{s}^{\prime\prime}}{z_{s}}\right)v_{k}=0.\label{eq12}
\end{equation}

We mentioned that the asymptotic limits    for  the Fourier modes
$v_k$ given by Eq.(\ref{eq12}) are  determined from the a small or
large scale limit. In this sense,
when the conformal time $\eta$ going to  $\eta\rightarrow -\infty$
corresponds to a small scale limit, in which the modes are inside
the horizon. On the contrary   when the conformal time
$\eta\rightarrow 0$ represents  to a large scale and then  the
modes to be outside the horizon. We also need to satisfy  the
Wronskian condition in which  $v_k^*\,v_k'-v_k\,v_k^{*'}=-i$, in
order to guarantee  the commutation relations from the quantum
theory\cite{noh}.

Under this formalism the curvature perturbation  in terms of
Fourier series is defined as

\be \mathcal{R}=\int
\frac{d^3\boldsymbol{k}}{(2\pi)^{3/2}}\mathcal{R}_{\boldsymbol{k}}(\eta)
e^{i\boldsymbol{k}\boldsymbol{x}}, \en and the vacuum expectation
value of $\mathcal{R}_{\boldsymbol{k}}$ is given by   $
\langle\mathcal{R}_{\boldsymbol{k}}\mathcal{R}_{\boldsymbol{l}}^*\rangle
= \frac{2\pi}{k^3}\mathcal{P}_R
\delta^3(\boldsymbol{k}-\boldsymbol{l}), $ where
$\mathcal{P}_R(k)$ corresponds to the power spectrum. In this way,
the power spectrum of the curvature perturbations is defined as

\be \mathcal{P}_R(k)=\frac{k^3}{2
\pi^2}\left|\frac{v_k}{z_s}\right|^2. \label{pp}\en

By knowing  the  power spectrum of the curvature perturbations,
then  the scalar index $n_s$ is given by
\begin{equation}
n_s-1=\frac{d\ln \mathcal{P}_R}{d \ln k}.\label{ns}
\end{equation}

On the other hand, it is well known that  the production of tensor
perturbation during the inflationary epoch would generate
gravitational waves. The framework to analyze the quantum
fluctuations in the gravitational field is similar to the case of
the scalar perturbation developed above. From metric, the tensor
perturbations are characterized by the tensor $h_{ij}$ and the
equation for the Fourier modes $u_k$ from the analysis
perturbative  is similar to Eq.(\ref{eq12}) and this equation is
given by
\begin{equation}
u_k''+\left(k^2-\frac{z_t''}{z_t}\right)\,u_k=0,\label{g}
\end{equation}
where now the new variable $z_t$ associated to the modes of the
tensor perturbations in the framework of the scalar tensor theory
is defined from the replacement of the scalar function
$Q_{s}\rightarrow Q_{t}=F$ \cite{shinji,noh} such that
\begin{equation}
z_{t}=a\,\sqrt{Q_t}=a\sqrt{F}.\label{eq13}
\end{equation}

In analogy  to  the  scalar perturbations, the power spectrum of
the tensor perturbations as function  of the modes $v_k$, is
denoted as
\begin{equation}
\mathcal{P}_g=\frac{2k^3}{\pi^2}\left|\frac{u_k}{z_t}\right|^2,
\end{equation}
where now the tensor spectral index $n_T$ is defined as $n_T=d\ln
\mathcal{P}_g/d\ln k$.

In the cosmological context,  a fundamental observational quantity
is related to the tensor to scalar ratio $r$, which is defined as
\begin{equation}
r=\frac{\mathcal{P}_g}{\mathcal{P}_R}.
\end{equation}

Observationally,  the BICEP2/Keck-Array collaboration
\cite{Array:2015xqh} published the upper bound on the
tensor-to-scalar ratio to be $r < 0.07$ ($95\%$ CL). Recently, the
Planck 2018 results\cite{Plancku} made it known that when the
likelihood is combined with B-mode polarization of
BICEP2/Keck-Array collaboration the ratio  $r < 0.064$ ($95\%$
CL).

The equations for the respective modes (\ref{eq12}) and
(\ref{g}), may be considered as a time independent
Schr$\ddot{o}$dinger equation of the type
$f''_k+\left(k^2-U\right)f_k=0$, where the effective potential as
function of the conformal time is defined as
$U=U(\eta)=z_{s}''/z_{s}$ for the scalar modes in which $f_k=v_k$
and $U=U(\eta)=z_{t}''/z_{t}$ for the tensor modes $f_k=u_k$,
respectively.

As we can observe from (\ref{eq12}) and  (\ref{g}), the evolution
 of the scalar and tensor cosmological  perturbations
during the inflationary epoch, are determined  by the functions
$z_s(\eta)$ and $z_t(\eta)$, however these  functions  at the same time are
related to the effective potential of cosmological perturbations $U(\eta)$.

A methodology used in the literature in order to find analytical solutions to the modes $v_k$ and $u_k$
is to consider
 an ``ansatz'' for the effective
potential of the cosmological perturbations $U(\eta)$, achieving
thus the reconstruction of the background variables, such as the
scalar potential $V(\phi)$ and the scale factor $a(t)$
analytically (or numerically). In particular in the framework of
the GR and considering  a standard scalar field analytical
solutions can be obtained  in order to reconstruct the background
variables (such as effective potential $V(\phi)$). In particular
 we have; de Sitter inflation in which the effective potential
of the scalar perturbation is given by $U(\eta)=\frac{2}{\eta^2}$
giving a potential $V(\phi)=$ constant  \cite{mukhanov}, the
expansion power law inflation
 ($a\sim t^p$ with $p>1$) where the effective potential of the scalar
perturbation results $U(\eta)=\frac{\nu^2-1/4}{\eta^2}$, where
$\nu$ is a constant defined as $\nu=3/2+1/(p-1)$ and the
reconstruction gives origin to an exponential potential associated
to scalar field\cite{st}. Also, considering   the ansatz for the
effective potential $ U(\eta)= C_1/\eta+C_2/\eta^2$,
 where $C_1$ and $C_2$ are constants  was studied in ref.\cite{martin} and the case in
 which $U(\eta)=0$ or equivalently $z_s=0$ was developed in \cite{Ea}.  In the
 context of a tachyon field, the effective potential for the scalar perturbations
 $U(\eta)=0$ together with  the ansatz $U(\eta)=C_0+C_1/\eta+C_2/\eta^2$, were analyzed in ref.\cite{Herrera:2015udk}. Here,
 we mentioned that
 the reconstruction of the scalar potential associated to the tachyonic field $V(\phi)$  for
 the ansatz
  $U(\eta)=C_0+C_1/\eta+C_2/\eta^2$ was obtained numerically.

In this sense, the evolution  of the  scalar and tensor
cosmological perturbations during the inflationary epoch are
completely determined from  the functions $z_{s}$ and $z_{t}$. In
the following,  we will develop a different  methodology  to
obtain the reconstruction of the background from the an  ansatz
for the potential of the cosmological perturbations $U(\eta)$.
Instead, we shall find a relation (differential equation)
 between the functions $z_s$ and $z_t$.  In fact, combining   Eqs.(\ref{eq4}) and
 (\ref{eq10}), we can eliminate the term $\omega\varphi^{\prime 2}$, then using
 the definitions of the functions $z_s$ and $z_t$ given by Eqs.(\ref{eq11}) and (\ref{eq13}),
 we obtain  that the relation between these variables is given by
\begin{equation}
z_{s}^{2}= 2 z_{t}^{2} \left[ 1 + \left(\frac{z_{t}}{z_{t}^{\prime}}\right)^{\prime}\,\right]. \label{eq14}
\end{equation}

In the following, we will study different solutions from the
relation given by Eq.(\ref{eq14}). From these solutions we will
find the background variables and its reconstruction  in the
Jordan frame as also the cosmological perturbations.

\section{Model I: $z_s\propto z_t$}

A  particular relation  between  the
variables $z_s$ and $z_t$ that satisfies  Eq.(\ref{eq14}), is given by
 $z_{s}=\gamma z_{t}$, in which  $\gamma$ denotes a
constant. We mention that this relation between $z_s$ and $z_t$ i.e., $z_s\propto z_t$ is equivalent to
consider the slow- roll approximation in the  Einstein gravity limit. In order to obtain this limit, we
have that the function
$F(\varphi)=\frac{1}{8 \pi G}=$ constant, and the parameter $\omega=1$, then Eq.(\ref{eq10})  can be written as
 $Q_{s}=\frac{\epsilon}{4\pi G}$, where $\epsilon$ is the
usual slow roll parameter defined as $\epsilon
=1-\frac{\mathcal{H}^\prime}{\mathcal{H}^{2}}$.  By assuming that
during the slow-roll regime, the parameter
 $\epsilon\simeq$  constant\cite{shinji}, and combining Eqs.(\ref{eq11}) and (\ref{eq13}), we
 get that
$z_{s}=\gamma z_{t}$ in which we identify  $\gamma=\sqrt{2\epsilon}$.

From the relation in which $z_s\propto z_t$,
we find  a particular solution to Eq.(\ref{eq14}) given by
\begin{equation}
z_{s}\propto z_{t}\propto\eta^q,\label{eq15}
\end{equation}
where  the power  $q=(\frac{\gamma^2}{2}-1 )^{-1}=$ constant
with $\gamma \neq \pm \sqrt{2}$. Note that in  the particular case in which
 $\gamma =\pm \sqrt{2}$, we have the solution  $z_{s}\propto z_{t}\propto e^{c \eta}$, in which $c=$ constant.

From the particular solution given by Eq.(\ref{eq15}) and considering  of the variable $z_t$ (see
Eq.(\ref{eq13})), we find that the arbitrary function $F$ becomes
\begin{equation}
F\propto\frac{\eta ^{2q}}{a^2}.\label{eq16}
\end{equation}

In this form, in order to reconstruct the function $F$ in terms of the scalar
field $\varphi$ from Eq.(\ref{eq16}), we have to give the scale factor $a$ in terms of the time $t$ (or $\eta$).

On the other hand, the cosmological perturbations in the case in
which $z_{s}\propto z_{t}$ can be determined, considering the
particular solution given by Eq.(\ref{eq15}). In this respect,
from Eqs.(\ref{eq12}) and (\ref{g}), the solutions for both modes
i.e., scalar and tensor modes in the large-scale limit are given
by \cite{shinji}
\begin{equation}
v_k(\eta)=u_k(\eta)\simeq -\frac{\sqrt{\pi \eta}}{2}\frac{i}{\pi}
\Gamma(\nu)\left(\frac{k\vert\eta\vert}{2}\right)^{-\nu},
\label{17}
\end{equation}
where the power $\nu$ is defined as $\nu\equiv \sqrt{q(q-1)+
1/4}$. Here, we mention that this solution for $v_k$ (or $u_k$) is
independent of the scale factor $a(t)$ how can it be seen of
Eq.(\ref{eq15}). From Eq.(\ref{17}), we find that the spectral
index and tensor-to-scalar ratio are given by \cite{shinji}
\begin{equation}
n_{s} =4-\sqrt{4q(q-1)+1}, \label{18}
\end{equation}
and
\begin{equation}
r=16( 1+1/q).\label{19}
\end{equation}
Note that these results are equivalent to those obtained during
power- law inflation in Einstein frame. The reason is because the
function $F$ connects  the conformal transformation between both
frames. Thus, from Eq.(\ref{eq16}) obtained under the particular
solution $z_s\propto z_t$ and considering the conformal
transformation
  $\hat{a}(\hat{t})=a(t)\sqrt{F}$ and $d \hat{t}=dt \sqrt{F}$, see \cite{shinji,kaiser},
 we have $\hat{a}(\hat{t}) \propto \hat{t}^{\frac{q}{1+q}}
 $ i.e., power law inflation in the Einstein frame.
 \footnotemark[3]\footnotetext[3]{Here the caret means quantity measured in Einstein frame.}.
In this point we emphasize that we have not assumed slow roll
approximation\footnotemark[4] and as we mentioned before the
explicit function for the scale factor  $a(t)$ does not  has been
specified yet.\footnotetext[4]{Indeed, starting from Eqs. (11) and
(\ref{eq13}) and one finds\cite{shinji}
$\frac{z_{s,t}^{\prime\prime}}{z_{s,t}}=\mathcal{H}^{2}\lbrace
(1+\delta_{s,t})(2-\epsilon+\delta_{s,t})+\frac{\delta_{s,t}^{\prime}}{\mathcal{H}})\rbrace$,
where $\delta_{s,t} =\frac{\dot{Q}_{s}}{2\mathcal{H}
Q}\equiv\delta_{s}$ for scalar perturbations and $\delta_{s,t}=
\frac{\dot{F}}{2\mathcal{H} F}\equiv\delta_{t}$ for tensor one. In
the slow roll approximation, $\epsilon$ and $\delta_{s,t}$ are
supposed to be constant during inflation stage, hence
$\frac{z_{s,t}^{\prime\prime}}{z_{s,t}}=\frac{\gamma_{s,t}}{\eta^{2}}$
with the constant $\gamma_{s,t} =\frac{(1+\delta_{s,t})(2-\epsilon
+\delta_{s,t})}{(1-\epsilon)^{2}}$. In contrast, from
Eqs.(11),(\ref{eq15}) and (\ref{eq16}),  we have $Q_{s}\propto F$,
hence $\delta_{s,t}= \frac{q}{\mathcal{H}\eta}-1$ and, as usual,
$\epsilon =1-\frac{\mathcal{H}^{\prime}}{\mathcal{H}^2}$. Thus, we
have
$\frac{z_{s,t}^{\prime\prime}}{z_{s,t}}=\frac{q(q-1)}{\eta^2}$. As
we can see, there is no  need to assume slow roll approximation.}


From the observational point of view, we have that from the Planck
data results, the scalar spectral index $n_s\simeq0.964$ and the
tensor to scalar ratio $r<0.1$\cite{Plancku}. From these
observational data, we find  that the parameter
$q=(\gamma^2/2-1)^{-1}$ takes two values; $q=-1.02$ and $q=2.02$,
by considering that the scalar spectral $n_s=0.94$. Thus,
evaluating the tensor to scalar ratio $r$ given by Eq.(\ref{19}),
we have that for the value $q=-1.02$ corresponds to $r\simeq0.3$
and for the value $q=2.02$ the parameter $r\simeq 24$. In this
form, we show that the model is not well supported by the
observational data, with which the solution $z_s\propto z_t$ does
not work.

Although the model $z_s\propto z_t$ does
not work, we consider
 an example  to reconstruct the background variables (scalar potential $V(\varphi)$ and coupling function $F(\varphi)$)for a specific inflationary
 model. In particular,
we consider the simplest situation in which the parameter $w=1$ or
induced gravity  and the inflationary expansion with the scale
factor  $a(t)\propto t^{p}$ ( or equivalently $  a\propto
\eta^{\frac{p}{1-p}})$ with $p>1$  (power-law inflation). Thus,
considering  Eq.(\ref{eq16}) we can integrate Eq. (\ref{eq4}),
obtaining that the scalar field $\varphi$ as function of the
conformal time results
\begin{equation}
\varphi(\eta)\propto \eta^{-\frac{p}{1-p}+q_1}, \label{20}
\end{equation}
with $q_1=$ constant such that $q_1\neq \frac{p}{1-p},-1$.

In this way, replacing Eq.(\ref{20}) in Eq.(\ref{eq16}), the reconstruction of the function $F(\varphi)$
during the power law inflation
takes the form
\begin{equation}
F(\varphi)= \alpha^{-2}\varphi^{2}\equiv \xi \varphi ^{2}, \label{21}
\end{equation}
where the constant $\alpha$ is defined as $\alpha
=\frac{c}{-\frac{p}{1-p}+q_1}$, in which
$c=\sqrt{-6(\frac{p}{1-p}-q_1)^{2} +
2q(1+q_1)}=\sqrt{\bigtriangleup}\neq 0$
 and $\bigtriangleup>0$ such that $c$
$\in\mathbb{R}$. Here, we can  identify the constant
$\alpha^{-2}=\xi$, wherewith corresponding to the theory of the
 Induced Gravity with the specific function $F(\varphi)$ given by Eq.(\ref{21})
 in which  $\xi>0$, see Refs.\cite{gi,gi2}.

By considering Eq.(\ref{potV}), we find  that the reconstruction of the effective potential $V(\varphi)$
 during the stage of power law inflation becomes

\begin{equation}
V (\varphi)\propto \varphi^{2\beta}, \label{22}
\end{equation}
where the power $\beta$ is defined as $\beta=
1-\frac{1}{q-p(1+q_1)}$.


\section{Model II: $ z_s=z_t\,f(\eta)$ }
In this section we develop a specific ansatz for the ratio between
variables $z_t$ and $z_s$ in terms of the conformal time $\eta$,
in order to obtain a solution to Eq.(\ref{eq14}). In this context,
we assume that the relation between the functions $z_s$ and $z_t$
is given by
\begin{equation}
z_s=z_t\,f(\eta),\label{zs}
 \end{equation}
where $f(\eta)$ is a function of
the conformal time $\eta$. Note that for the model I studied
previously, we have the specific case in which   the function
$f(\eta)=$ constant, such that $z_t\propto z_s$.

By replacing Eq.(\ref{zs}) in Eq.(\ref{eq14}) we find that the
function $f(\eta)$ is related with the tensor variable $z_t$ as
\begin{equation}
f(\eta)^2=2\left[1+\left(\frac{z_t}{z_t'}\right)'\right]=2\left[1+\left(\frac{1}{(\ln z_t)'}\right)'\,\right].\label{f1}
 \end{equation}
In this form, we find that the general solution for the  tensor function $z_t$ in terms of the $f(\eta)$ can be written as
 \begin{equation}
z_t(\eta)=\exp\left[\int\left(\int\left[\frac{f(\eta)^2}{2}-1\right]d\eta\right)^{-1}d\eta
\right].\label{gs}
\end{equation}
Thus, we can give a specific  function or ansatz for $f$ is terms
of the conformal time, in order to obtain the tensor function
$z_t$.

An interesting situation occurs when we choose that the function
$f(\eta)$ is given by the ansatz
\begin{equation}
f(\eta)=\sqrt{2}\,\left[1+\frac{\beta_1}{(\beta_1-\alpha_1|\eta|)^2}\right]^{1/2},\label{f2a}
\end{equation}
where $\beta_1$ and $\alpha_1$ are constants. The constant $\beta_1$ is a
dimensionless constant and $\alpha_1$ has dimensions of mass.

In this sense, replacing Eq.(\ref{f2a}) in  (\ref{gs}), we find
that a particular solution for the tensor function $z_t$ can be
written as
\begin{equation}
z_{t}(\eta)\propto |\eta|\,^{\beta_1}e\,^{-\alpha_1|\eta|}.
\label{28a}
\end{equation}

Thus, from solution (\ref{28a}), we obtain that  the effective
potential $U(\eta)$ of the tensor perturbation becomes
\begin{equation}
U(\eta)=\frac{z_{t}^{\prime \prime }}{z_{t}}=C_{0}^{t}+
\frac{C_{1}^{t} }{\mid\eta\mid }+\frac{C_{2}^{t}}{\eta ^{2}},
\label{23a}
\end{equation}
where the constants $C_{0}^{t}$, $C_{1}^{t}$ and $C_{2}^{t}$ are
defined as
$$
C_{0}^{t}=\alpha_1^2,\,\,\,\,C_{1}^{t}=-2\alpha_1\,\beta_1,\,\,\,\,\mbox{and
}\,\,\,\,C_{2}^{t}=\beta_1(\beta_1-1),
$$
respectively.  Then, the constant $C_0^t$ has dimensions of mass
squared, $C_1^t$ has dimensions of mass and $C_2^t$ is a
dimensionless constant. Also, we note that the constant
$C_{1}^{t}$ depends on the constants  $C_{0}^{t}$ and $C_{2}^{t}$,
respectively.

 From Eq.(\ref{23a})  we note  that a general solution  for the tensor variable $z_t$  can
 be written as
  \begin{equation}
z_{t}(\bar{\eta})= A_{1}^{t} W_{\alpha_{t},\beta}(\bar{\eta})+
A_{2}^{t} W_{-\alpha_{t},\beta}(-\bar{\eta}),  \label{24aa}
\end{equation}
where the function $W_{\pm\alpha_{t},\beta}$ corresponds to the
 Whittaker function \cite{Wi} and $A_1^t$ together with  $A_2^t$ are two integration constants.
 Note that for the solution (\ref{24aa}),
 we have used  the change of the  variables given by
$$
\bar{\eta}=2\sqrt{C_{0}^{t}}\mid\eta\mid=2\alpha_1\mid\eta\mid,\,
\,\alpha_{t}=-\frac{C_{1}^{t}}{2\sqrt{C_{0}^{t}}}=\beta_1,\,\,\mbox{and
}\,\,\,\,\beta^{2}=\frac{1}{4} + C_{2}^{t}=\frac{1}{4} + \beta_1(\beta_1-1),
$$
respectively.

In this sense, we can recognize that the particular  solution (\ref{28a}) corresponds
to the  large scale limit ($\mid\eta\mid \longmapsto 0$), of
the  Whittaker function $W_{ \alpha_{t},\beta}$ defined as  \cite{Wi}
\begin{equation}
z_{t}(\eta)\rightarrow
B_{t}(2\sqrt{C_{0}^{t}}\mid\eta\mid)^{\frac{1}{2}-\beta}
e^{-\sqrt{C_{0}^{t}}\mid\eta\mid}\propto
|\eta|\,^{\beta_1}e\,^{-\alpha_1|\eta|}=C_0\,|\eta|\,^{\beta_1}e\,^{-\alpha_1|\eta|},
\label{28aa}
\end{equation}
where we can recognize to $\beta_1=\frac{1}{2}-\beta$ and the constant
 $C_0=B_{t}(2\sqrt{C_{0}^{t}}
 )^{\frac{1}{2}-\beta}$.

In order to take the large scale limit of the Whittaker function
$W_{ \alpha_{t},\beta}$ given by Eq.(\ref{24aa}),  we have defined
$B_{t}\equiv
A_{1}^{t}\frac{\Gamma(2\beta)}{\Gamma(\frac{1}{2}+\beta
-\alpha_{t})}=A_{1}^{t}\frac{\Gamma(2\beta)}{\Gamma(\frac{1}{2}+\beta
-\beta_{1})}=C_0\,(2\alpha_1
 )^{\beta-\frac{1}{2}}$ where $\Gamma$ corresponds to the Gamma function and we also choose $A_{2}^{t}=0$ in the general solution given
by  Eq.(\ref{24aa}) without loss of generality.

Thus, in relation to the spectrum  of gravity waves, we note that
the equation for the tensor-modes $u_k$ combining Eqs.(\ref{g})
and (\ref{23a}) can be written as
\begin{equation}
\frac{d^2u_k(\eta)}{d\eta^2}+\left(k^2-C_0^t-\frac{C_1^t}{\mid\eta\mid}-\frac{C_2^t}{\eta^2}\right)u_k(\eta)=0.\label{UK}
\end{equation}

From Eq.(\ref{UK}), we find that the solution for the Fourier
modes $u_k$ can be written as
\begin{equation}
u_{k}(\tilde{\eta})= B_{1}
W_{\tilde{\alpha_{t}},\beta}(\tilde{\eta})+ B_{2}
W_{-\tilde{\alpha_{t}},\beta}(-\tilde{\eta}), \label{41}
\end{equation}
where the conformal time $\tilde{\eta}$ is defined as
$\tilde{\eta}\equiv 2 i k_{eff}^{t}\mid\eta\mid$,
$k_{eff}^{t}\equiv \sqrt{k^{2}- C_{0}^{t}}$ and
$\tilde{\alpha_{t}}\equiv \frac{C_{1}^{t}\,\,i}{2k_{eff}^{t}}$.
Also, the constants $B_{1}$ and $B_{2}$ correspond to integration
constants and these are fixed by initial conditions. In particular
for de Sitter universe in the frame of Einstein (where $F=1$) the
solution for the tensor-modes $u_k=(1/\eta)[B_1(k\eta
\cos[k\eta]-\sin[k\eta])+B_2 (k\eta \cos[k\eta]+\sin[k\eta])]$
where $C_0^t=0$, $C_1^t=0$ and $C_2^t=2$, since the scale factor
$a\propto 1/\mid \eta\mid$.

From Eqs.(\ref{24aa}) and (\ref{41}), the power spectrum of the
tensor perturbation becomes
\begin{equation}
{\cal{P}}_g=\frac{2k^3}{\pi^2}\left|\frac{u_k}{z_t}\right|^2=\frac{2k^3}{(\pi\,A_1^t)^2}\frac{\mid
B_{1} W_{\tilde{\alpha_{t}},\beta}(\tilde{\eta})+ B_{2}
W_{-\tilde{\alpha_{t}},\beta}(-\tilde{\eta})\mid^2}{\mid
W_{\alpha_t,\beta}(\bar{\eta})\mid ^2}.\label{w1}
\end{equation}

In order to determine the constants $B_1$ and $B_2$, it is
necessary  to consider the quantities associated to the tensor
modes $u_k(\eta_i)$ and $u'_k(\eta_i)$
 at initial conformal time $\eta_i$. Following Ref.\cite{mukhanov},
we can consider that the initial conditions for the modes
$u_k(\eta_i)$ and $u'_k(\eta_i)$ are given by
\begin{equation}
u_k(\eta_i)=u_{k_{\eta_i}}=\left[k^2-C_0^t-\frac{C_1^t}{\mid\eta_i\mid}-\frac{C_2^t}{\eta_i^2}\right]^{-1/4},\,\,\,
\,\,\mbox{and}\,\,\,\,\,u'_k(\eta_i)=\frac{i}{u_{k_{\eta_i}}}.\label{cond3}
\end{equation}

In this context, it is possible  to obtain the values of the
integration constants $B_1$ and $B_2$, considering   the
asymptotic conduct of the Whittaker function
$W_{\tilde{\alpha_{t}},\beta}(\tilde{\eta})$ in the small-limit
($\mid\eta\mid\to \infty$) in which
$W_{\tilde{\alpha_{t}},\beta}(\tilde{\eta})\to
\mid\tilde\eta\mid^{\tilde{\alpha_{t}}}\exp(-ik_{eff}\mid\eta\mid)$
and together with the initial conditions given by Eq.(\ref{cond3})
we have
\begin{equation}
B_{1}= \frac{\tilde{\alpha_{t}}}{2\sqrt{ k_{eff}^{t}}}
\frac{(2k_{eff}^{t}\mid\eta_{i}\mid
)^{-\frac{iC_{1}^{t}}{2k_{eff}^{t}}}}{
(\tilde{\alpha_{t}}-ik_{eff}^{t}\mid\eta_{i}\mid )}
e^{ik_{eff}^{t}\mid\eta_{i}\mid+\frac{\pi
C_{1}^{t}}{4k_{eff}^{t}}}, \label{42}
\end{equation}
and
\begin{equation}
B_{2}= \frac{1}{2\sqrt{k_{eff}^{t}}}
\frac{(\tilde{\alpha_{t}}-2ik_{eff}^{t}\mid\eta_{i}\mid)}{
(\tilde{\alpha_{t}}-ik_{eff}^{t}\mid\eta_{i}\mid)}(-2k_{eff}^{t}\mid\eta_{i}\mid
)^{\frac{iC_{1}^{t}}{2k_{eff}^{t}}})e^{-(ik_{eff}^{t}\mid\eta_{i}\mid+\frac{\pi
C_{1}^{t}}{4k_{eff}^{t}})}, \label{43}
\end{equation}
respectively. Here, we have  considered that in the limit in which
$\mid\eta\mid\to \infty$ (or small-scale), the quantity
$(k^2-C_0^t-C_1^t/\mid\eta_i\mid-C_2^t/\eta_i^2)\simeq(k^2-C_0^t)>0$.

In this way, the power  spectrum of the tensor perturbation in the
large scale limit (or equivalently  $\mid\eta\mid\to 0$) from
Eq.(\ref{w1}) becomes

$$
{\cal{P}}_{g}(k)= \frac{2k^{3}}{\pi^{2}A_{1}^{t2}} \mid \Gamma(1/2
+\beta
-\alpha_{t})\mid^{2}\left(\frac{k_{eff}^{t2}}{C_{0}^{t}}\right)^{1/2-\beta}\Bigl[
\frac{\vert B_{1}\vert^{2}} {\mid \Gamma(1/2 +\beta
-\tilde{\alpha_{t}})\mid^{2}} +
$$
\begin{eqnarray}
 \frac{\vert B_{2}\vert^{2}}
{\mid \Gamma(1/2 +\beta +\tilde{\alpha_{t}})\mid^{2}}
+(-1)^{1/2-\beta} \mbox{Re}\left(\frac{2B_{1}
B_{2}^{\ast}}{\Gamma(1/2 +\beta
-\tilde{\alpha_{t}})\Gamma^{\ast}(1/2 +\beta
+\tilde{\alpha_{t}})}\right)\Bigr].\label{45}
\end{eqnarray}


On the other hand, we find that the scalar function $z_s$ in the
large scale limit, from Eqs.(\ref{zs}), (\ref{f2a}) and
(\ref{28aa}) can be written as
\begin{equation}
z_s(\eta)\simeq \sqrt{2}\,B_t\,(2\alpha_1)^{\beta_1}
|\eta|\,^{\beta_1}e\,^{-\alpha_1|\eta|}\,\,\left[1+\frac{\beta_1}{(\beta_1-\alpha_1|\eta|)^2}\right]^{1/2},
\end{equation}
or equivalently
\begin{equation}
z_{s}(\eta)=\frac{\sqrt{2}B_{t}(2\sqrt{C_{0}^{t}}\mid\eta\mid)^{\frac{1}{2}-\beta}
e^{-\sqrt{C_{0}^{t}}\mid\eta\mid}}{(1-2\beta)(2\sqrt{C_{0}^{t}}\mid\eta\mid)^{-1}-1}\sqrt{1+(1-2\beta)
(2\sqrt{C_{0}^{t}}\mid\eta\mid)^{-2}\lbrace3-2\beta-4\sqrt{C_{0}^{t}}\mid\eta\mid\rbrace}.
\label{29}
\end{equation}
Assuming the large scale limit in which $\mid\eta\mid \longmapsto
0$, we can expand  in power series the expression $z_{s}^{\prime
\prime }/z_{s}$ or the effective potential associated to the
scalar function  in which the  first order perturbed equation
(\ref{eq12}) for the scalar modes $v_{k}$ can be written as





\begin{equation}
v_{k}^{\prime\prime}(\eta) +
\left(k^{2}-C_{0}^{s}-
\frac{C_{1}^{s}}{\mid\eta\mid }-\frac{C_{2}^{s}}{\eta ^{2}}\right)v_{k}(\eta)\simeq 0,\label{eq12b}
\end{equation}
where the constants $C_{i}^{s}$ are functions of the $C_{i}^{t}$,
such that

\begin{equation}
C_{0}^{s}= \frac{81-160\beta+152 \beta{^2}-64 \beta^{3}+16
\beta^{4}}{(3-8\beta + 4 \beta^{2})^{2}}C_{0}^{t}\,,  \label{31}
\end{equation}
\begin{equation}
C_{1}^{s}= \frac{-1-8\beta+4\beta{^2}}{(-3+2\beta
)}\sqrt{C_{0}^{t}},\,\,\,\,\;\;\;\mbox{and}\,\,\,\,\,\;\;\;\;\;C_{2}^{s}=
-\frac{1}{4}+ \beta^{2}= C_{2}^{t}.\label{32}
\end{equation}

Redefining the conformal time $\eta$ together with the constants
$C_0^s$, $C_1^s$ and $C_2^s$ (or $\beta$) such that;
$\tilde{\eta}\equiv 2 i k_{eff}^{s}\mid\eta\mid$ in which
$k_{eff}^{s}\equiv \sqrt{k^{2}- C_{0}^{s}}>0$,
$\tilde{\alpha_{s}}\equiv \frac{C_{1}^{s}i}{2k_{eff}^{s}}$ and
$C_2^s$ (see Eq.(\ref{32})),  we have that the general solution of
Eq.(\ref{eq12b}) for the Fourier modes $v_{k}$ is given by
\begin{equation}
v_k(\tilde{\eta})=A_0\,W_{\tilde{\alpha_{s}},\beta}(\tilde{\eta})+A_1\,W_{-\tilde{\alpha_{s}},\beta}(-\tilde{\eta}),
\end{equation}
in which $A_0$ and $A_1$ are two integration constants and
$W_{\tilde{\alpha_{s}},\beta}$ denotes a new Whittaker function.

On the other hand, the constants $A_0$ and $A_1$ can be determined
considering that in the small scale limit in which $\mid\eta\mid\to \infty$, the Fourier modes
$v_k\to\frac{1}{\sqrt{2k_{eff}^s}}\exp[-ik_{eff}^s\tilde{\eta}]$ together with
the Wronskian condition. Thus, we obtain that the integration constant $A_0=\frac{e^{\frac{ \pi C_{1}^{s}}{4
k
_{eff}}}}{\sqrt{2k_{eff}^{s}}}$ and the constant
$A_1=0$, in order to satisfy these conditions.
Thus, we find that the Fourier modes $v_{k}$ can be written as
\begin{equation}
v_{\kappa}(\tilde{\eta})=\frac{e^{\frac{ \pi C_{1}^{s}}{4
k
_{eff}}}}{\sqrt{2k_{eff}^{s}}}W_{\tilde{\alpha_{s}},\beta}(\tilde{\eta}).\label{35}
\end{equation}

In order to obtain the scalar power spectrum of the curvature perturbations, we
need to
consider the growing modes of the $v_k$. These large-modes   are obtained
at large scale limit of the function $v_k(\mid\eta\mid\to 0)$, archiving thus that the quantum fluctuations are
frozen outside horizon. In
this sense, considering that the behavior of the Whittaker function when $\mid\tilde{\eta}\mid\to 0$
is given by $$
W_{\tilde{\alpha_{s}},\beta}(\tilde{\eta})\rightarrow
\frac{\Gamma(2\beta)}{\Gamma(s)}
\tilde{\eta}^{1/2-\beta}e^{-\tilde{\eta}/2} \,\,\,\,\,\mbox{with}\,\,\,
\, s\equiv 1/2 + \beta- \tilde{\alpha_{s}}.$$
In this way, from Eq.(\ref{35}) we get that the scalar modes $v_k$
for large scale as function of the conformal time $\eta$ becomes
\begin{equation}
v_k(\eta)=\frac{e^{\frac{ \pi C_{1}^{s}}{4
k
_{eff}}}}{\sqrt{2k_{eff}^{s}}}\,\frac{\Gamma(2\beta)}{\Gamma(s)}
[2ik_{eff}^s\eta]^{1/2-\beta}e^{-i k_{eff}^s\eta}.\label{51}
\end{equation}

In order to obtain the scalar power spectrum ${\cal{P_R}}(k)$, we
consider that the spectrum is evaluated when the wavelength of the
perturbation crosses the Hubble radius i.e., $k^2 =
z_s^{\prime\prime}/z_s$, (see e.g., the review the inflation in
Ref. \cite{Tsujikawa:2003jp}). From Eq.(\ref{eq12b}) we find that
$z_s^{\prime\prime}/z_s\simeq C_0^s + C_1^s/\eta + C_2^s/\eta^2$,
with which we have that the conformal time when the wavelength of
the perturbation crosses the Hubble radius is given by
$-\eta(k)=-\eta_\ast=[-C_{1}^{s}+\sqrt{C_{1}^{s\,2}-4C_{0}^{s}C_{2}^{s}+4C_{2}^{s}k^2}\,]/[2(k^2-C_{0}^{s})]$.
We note that in the special case of de Sitter inflation $C_0^s =
C_1^s = 0$, then $-k\eta \simeq 1$
 and $k \simeq aH$.

By combining  Eqs.(\ref{pp}) and (\ref{51}), we find that scalar
power spectrum ${\cal{P_R}}(k)$
 becomes
 \begin{equation}
{\cal{P_{R}}}(k)=\left(\frac{k^{3/2}}{2 \pi\,C_0}\right)^2
\,e^{2\alpha_{1} \mid\eta_*\mid +\frac{\pi C_{1}^{s}}{2
k_{eff}^{s}}}\left[1+\frac{\beta_{1}}
{(\beta_{1}-\alpha_{1}\mid\eta_*\mid)^2}\right]^{-1}
\frac{\Gamma^{2} (2\beta)} {\Gamma (s^{\ast}) \Gamma (s)} (2
k_{eff}^{s} )^{-2\beta} ,\label{37}
\end{equation}
and as before the quantity  $\Gamma$ denotes the Gamma function.
Now, combining Eqs.(\ref{ns}) and (\ref{37}), we obtain that the
scalar spectrum index $n_s$ can be written as


\begin{displaymath}
n_{s}(k)=1 + \frac{d \ln \cal{P_{R}}}{d \ln k}= 4-\frac{k^2}{k_{eff}^{s}}\left[ \frac{2\beta}{k_{eff}^{s}} +
\frac{C_{1}^{s}}{ 2 k_{eff}^{s2}}(\pi + i [\Psi_{0}(s)- \Psi_{0}(s^{\ast}) ])\right] \,\,+
\end{displaymath}
\begin{equation}
4k^2 \alpha_{1}(k^2-C_{0}^{s})^{-1}\ \left[\mid \eta_{\ast}\mid +
\frac{C_{2}^{s}}{C_{1}^{s}-2\mid \eta_{\ast}\mid (k^2-C_{0}^{s}) }
\right] \left[\frac{\beta_{1}(\beta_{1}-\alpha_{1}\mid \eta_*\mid
)^{-1}} {\beta_{1}+ (\beta_{1}-\alpha_{1}\mid \eta_{\ast}\mid
)^{-2} }-1\right],\label{ns3}
\end{equation}
where the quantity $\Psi_0(s)$ corresponds to   the Polygamma function.

\begin{figure}[th]
{{\vspace{0.0 cm}\includegraphics[width=5 in,height=3 in,angle=0,clip=true]{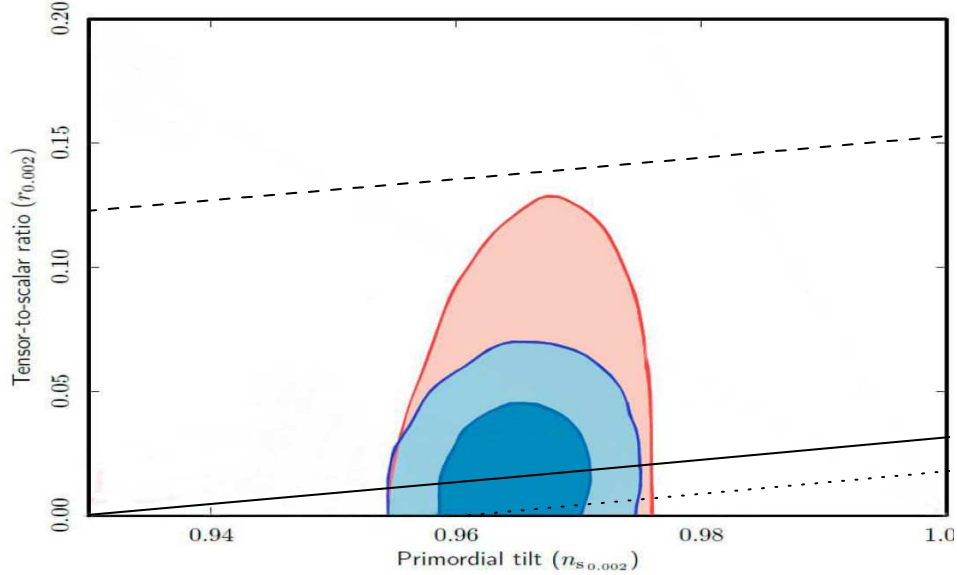}}}
{\vspace{0.0 cm}\caption{  The panel shows the tensor-to-scalar ratio $r$ as a
function of the scalar spectral index $n_s$,  for
three different values of the parameter $A_1^t$.  In this panel we have
considered the two-marginalized constraints joint 68$\%$ and 95$\%$ CL at $k=0.002$ Mpc$^{-1}$ from the Planck 2018 results
\cite{Plancku}.
In the plot
 the
dashed, solid and dotted lines correspond to the values of parameter
$A_1^t=0.99$, $A_1^t=0.999$ and $A_1^t=0.1$, respectively.
 \label{fig1}}}
\end{figure}


In Fig.\ref{fig1} we show the tensor to scalar  ratio $r$ versus the scalar spectral index $n_s$,
for three different values
of the parameter  $A_1^t$ for the model II, by assuming  the ansatz given by  Eq.(\ref{f2a})
 for the function $f(\eta)$. In this panel we use the two-marginalized constraints for
the  relation $r=r(n_s)$ (at 68$\%$ and 95$\%$ CL at  $k=0.002$
Mpc$^{-1}$ ) from the new Planck results\cite{Plancku}. By
considering  Eqs.(\ref{45}), (\ref{37}) and (\ref{ns3}), we
numerically obtain the parametric plot for tensor to scalar ratio
in terms of the scalar spectral index or the consistency relation
$r=r(n_s)$ for three values of the parameter $A_1^t$ (integration
constant, see Eq.(\ref{24aa})). In this plot, we have considered
that  the dashed, solid and dotted lines correspond to the values
$A_1^t=0.99$, $A_1^t=0.999$ and $A_1^t=0.1$, respectively.  On the
other hand,
 from Eqs.(\ref{37}) and (\ref{ns3}) we can note that the scalar power  spectrum  ${\cal{P}_R}$
 depends of the parameters
 ${\cal{P}_R}={\cal{P}_R}(k,\beta, C_0^t,A_1^t)$ and the scalar spectral index  $n_s=n_s(k,\beta,C_0^t)$.
In this way,
we numerically
find from these relations that the values of the
parameters $\beta=-1.35997$ (or $\beta_1=1.85997 $ since $\beta_1=1/2-\beta$) and
$C_0^t=3.947\times10^{-3}$ (or $\alpha_1\simeq 0.063$ since $\alpha_1=\sqrt{C_0^t}$) correspond to the parameter $A_1^t=0.1$
(dotted line). Here we
have used  the values of ${\cal{P}_R}=2.2\times 10^{-9}$ and
$n_s=0.964$ at $k=0.002$Mpc$^{-1}$. Also, we find that the tensor to scalar ratio
for these values at $k=0.002$Mpc$^{-1}$ is given by $r(k=0.002)=0.0015$. Analogously, we numerically
find  that for the value $A_1^t=0.999$, we have the parameters
$\beta=-1.35998$ and $C_0^t=3.963\times10^{-3}$ (solid line). Here we obtain that the tensor
to scalar ratio at  $k=0.002$Mpc$^{-1}$ corresponds to $r(k=0.002)=0.0016$.
For the situation in which $A_1^t=0.99$ (dashed line), we numerically find
the values of the parameters $\beta=-1.3601$, $C_0^t=4.110\times10^{-3}$  and the ratio
 $r(k=0.002)=0.138$.
Here, we have  considered  that the initial conformal time
$\mid\eta_i\mid=400.000$, since the tensor to scalar ratio $r$
depends of $r=r(k,\beta, C_0^t,A_1^t,\eta_i)$. In this form, from
Fig.\ref{fig1} we note  that the value of the
 integration constant  $A_1^t\simeq 0.1$ since it is well corroborated by Planck
2018 results. On the other hand, for  values of the constant $A_1^t<0.1$,
we find that the ratio $r>0.1$ and then the model II is disfavored from Planck data.
Also, we mention that for values of the integration constant $A_1^t>0.1$ the
tensor to scalar ratio $r$ becomes negative and the model is excluded from
observational data.
 Thus,
    from Fig.1, we observe that for the values
of the parameters $A_1^t\simeq 0.1$, $\beta\simeq-1.4$ and
$C_0^t\simeq 10^{-3}$, the model II describe for the ansatz given
by  Eq.(\ref{f2a}), is well corroborated by Planck 2018 data,
however we note that the parameter-space is very small. Also, we
note that from Eqs.(\ref{31}) and (\ref{32}), we can  obtain the
values $C_0^S\simeq0.007$, $C_1^S\simeq-0.2 $ and $C_2^S\simeq
1.6$, respectively. This suggests that parameter-space for our
model II is similar to de Sitter Universe in which $C_0^S=0$,
$C_1^S=0$ and $C_2^S=2$.


 On the other hand, from Eqs.(\ref{eq13}), (\ref{28a}) and (\ref{24aa}) we
find that the arbitrary function $F$ in terms  of the time $\eta$
can be written as
\begin{equation}
F(\eta)=\frac{z_t^2}{a^2}=
a^{-2}A_{1}^t\,^{2}\,W_{\alpha_{t},\beta}(\eta)^{2}\propto
a^{-2}\,\mid\eta\mid^{2\beta_1}\,e^{-2\alpha_1\mid\eta\mid} .
\label{46}
\end{equation}
Here we have considered the large scale limit of the Whittaker function
$W_{\alpha_{t},\beta}$.

In order to build the effective potential, we find that combining
Eqs.(\ref{potV}) and (\ref{eq13}), the potential can be rewritten
as function of the variable $z_t$ as
\begin{equation}
V= a^{-4}\lbrace \,z_{t}^{\prime\, 2} +
z_{t}\,z_{t}^{\prime\prime}\,\rbrace.    \label{47}
\end{equation}
In this way, we find the effective potential in terms of the
conformal time $\eta$ results

\begin{equation}
V(\eta)=\frac{F}{a^2}
\left[2x^2-(\mathcal{H}+x)^2-\frac{\beta_1}{\eta^2}\right],\label{48}
\end{equation}
where the quantity $x$ is defined as
$x=\frac{\beta_1}{|\eta|}-\alpha_1$ and the function $F$ is given
by Eq.(\ref{46}). Here we have also used that
$F^{\prime}=2F[x-\mathcal{H}]$.

From Eq.(\ref{eq4}) (or equivalently of Eq.(\ref{eq10})) we obtain
that the relation between the scalar field and the conformal time
can be found from the differential equation given by
\begin{equation}
\varphi^{\prime\,2}(\eta)=\frac{2F}{\omega}\left[x^2-3(x-\mathcal{H})^2+\frac{\beta_1}{\eta^2}\right].\label{ff}
\end{equation}

On the other hand, in order to reconstruct the background, in
particular the effective potential $V(\phi)$ and the scale factor
$a(\eta)$, we will consider a simple example in which the coupling
function $F(\varphi)\propto\varphi^2=B_0\varphi^2$ and also the
parameter $\omega=$const. In this case, we have that
$F^{\prime}=2B_0\varphi\varphi^{\prime}=2F(x-\mathcal{H})=2B_0\varphi^2(x-\mathcal{H})$
and combining with Eq.(\ref{ff}) we get
\begin{equation}
\frac{\varphi'}{\varphi}=D_0\,\left[x^2+\frac{\beta_1}{\eta^2}\right]^{1/2},\,\,\,\,\,\mbox{where}\,\,\,\,D_0=\sqrt{\frac{2B_0}{6B_0+\omega}}.\label{ss}
\end{equation}
Here we have considered the positive sign for simplicity. In this
form, we find that the solution of the Eq.(\ref{ss}) can be
written as
\begin{equation}
\varphi(\eta)=\varphi_0\,\,
e^{D_0y}\left(\frac{|\eta|}{2b_1+b_2|\eta|+2\sqrt{b_1}\,y}\right)^{D_0\sqrt{b_1}}\left(b_2+2\,|\eta|
b_3+2\sqrt{b_3}\,y\right)^{-D_0\beta_1},\label{ff2}
\end{equation}
where the quantity $y$ is defined as
 $y=y(\eta)=\sqrt{b_1+|\eta| b_2+\eta^2\,b_3)}$, $\varphi_0$ denotes
 an
 integration constant and the constants $b_1$, $b_2$ and $b_3$ are
 defined as
$$
b_1=\beta_1(\beta_1+1),\,\,\,\,\,\,
b_2=-2\beta_1\alpha_1,\,\,\,\,\,\, b_3=\alpha_1^2,
$$
respectively.

By combining Eqs.(\ref{46}) and (\ref{ff2}) we obtain that the reconstruction for the scale factor as a function of the
conformal time $\eta$ is given by
\begin{equation}
a(\eta)\propto
e^{-(\alpha_1|\eta|+D_0y)}\left(2b_1+b_2|\eta|+2\sqrt{b_1}\,y\right)
^{D_0\sqrt{b_1}}\left(b_2+2\,|\eta|
b_3+2\sqrt{b_3}\,y\right)^{D_0\beta_1}\,|\eta|
^{\beta_1-D_0\sqrt{b_1}}.\label{aas}
\end{equation}

\begin{figure}[th]
{{\vspace{0.0 cm}\includegraphics[width=3.0
in,angle=0,clip=true]{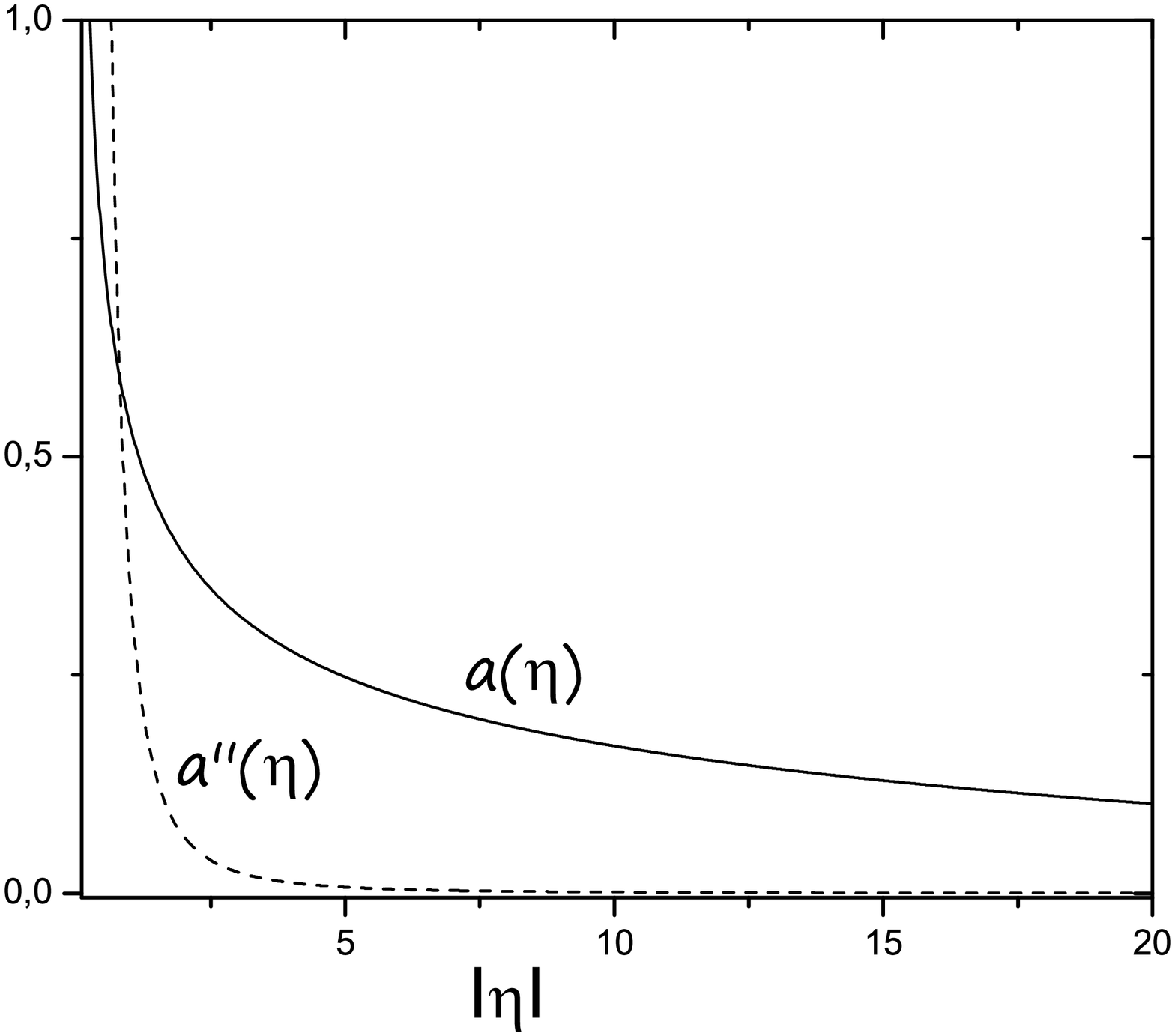}}} {{\vspace{-1.5
cm}{\includegraphics[width=3.0in,angle=0,clip=true]{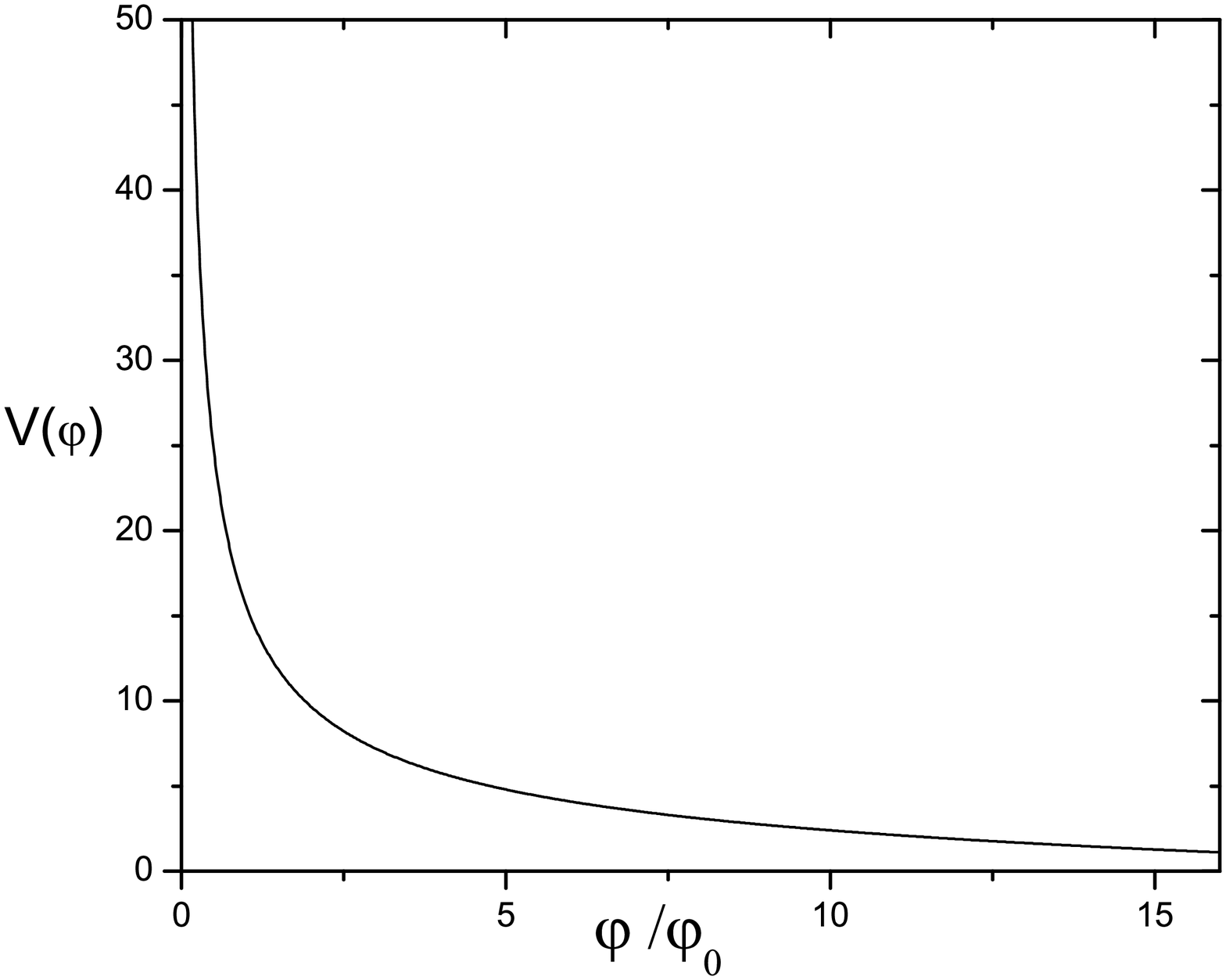}}}}
{\vspace{1.0 cm}\caption{  In the left panel we show the scale
factor $a(\eta)$ (solid line) and its acceleration $a''(\eta)$
(dashed line) as a function of the conformal time $\eta$. In the
right panel we show
 the effective potential $V(\varphi)$ in terms of the scalar field. In both
 panels we have considered $D_0=1$, $C_0^t=3.947\times 10^{-3}$ and
 $\beta=-1.35997$.
 \label{fig2}}}
\end{figure}

In Fig.\ref{fig2}, we show the evolution of the scale factor
$a(\eta)$ panel) and its acceleration  $a''(\eta)$ versus the
conformal time $\eta$(left panel). In the right panel we show the
effective potential in terms of the scalar field $\varphi$.
 In these panels, we have
used  the parameters  $D_0=1$,  $C_0^t=3.947\times 10^{-3}$ and
 $\beta=-1.35997$, respectively.

 For the plot of the scale factor in the left panel, we consider the analytical
reconstruction given by  Eq.(\ref{aas}) for  $a(\eta)$ (see solid
line) and checking its acceleration $a''(\eta)>0$ (dashed line).
On the other hand,
 in order to build the effective potential as a function of
the scalar field i.e., $V(\varphi)$, we have considered
Eq.(\ref{48}) and the solution $\varphi(\eta)$ given by
Eq.(\ref{ff2}). Thus, we can reconstruct the effective potential
as a function of the scalar field $\varphi$ considering a
parametric plot from $V(\eta)$ and $\varphi(\eta)$, respectively
(see right panel).



\section{Conclusions}

In this article, we have studied an approach to
exact solutions of cosmological perturbations in the context of a generalized
scalar tensor theory for two models characterized for the  relation between the variables of the modes
of the tensor ($z_t$) and scalar ($z_s$) perturbations, respectively.

 For both models, we have obtained analytical solutions
 for the corresponding  power spectrum of
 the curvature perturbations, scalar spectral index, the
 power spectrum of the tensor perturbations, and the tensor to scalar ratio
  together with the background variables.

For the model I we have considered that the relation between the
modes is given by $z_s\propto z_t$. For this model we have found a
particular solution in which $z_s\propto z_t\propto \eta^q$, where
the parameter $q$ has relation with the proportionality constant.
For this model we have obtained that this model is not well
supported by the Planck data, since the tensor to scalar ratio
$r>0.1$. This suggests that the model I in which $z_s\propto z_t$
does not work. However, although the model $z_s\propto z_t$ is not
well agreement from the observational data, we have considered
 an example  to reconstruct the background variables (scalar potential $V(\varphi)$ and coupling
 function
 $F(\varphi)$), by assuming the power-law inflation and the parameter $\omega=1$. In
 this context,  we have found that the coupling function $F(\varphi)$ is given by $F(\varphi)\propto
 \varphi^2$ and the effective potential corresponds to the power-law potential
  in which $V(\varphi)\propto\varphi^{2\beta}$ (see
 Eq.(\ref{22})).

For the model II we have assumed a more general relation between
the variables $z_s$ and $z_t$ given by Eq.(\ref{zs}). In order to
obtain analytical solution to the cosmological perturbations, we
have considered an interesting ansatz-function $f(\eta)$ given by
 Eq.(\ref{f2a}). From this ansatz on $f(\eta)$, we have found that the Fourier modes of the scalar
and tensor perturbations can be written in terms of the Whittaker
functions. By assuming  the large and small limits for each one of
the Fourier modes, we have found analytical expressions for the
scalar and tensor power spectral as a function of the  wave number
$k$. From these expressions for the amplitudes, we have obtained
the constraints on the different  parameters  considering  the new
Planck 2018 results. In this context, we have used the consistency
relation $r=r(n_s)$, and we have noted that our parameter-space is
very reduced. In this sense, we have observed from the
plane-trajectories $r=r(n_s)$ the model II is well supported by
Planck results only if the parameters in our model are
$C_0^S\simeq0.007$, $C_1^S\simeq-0.2 $ and $C_2^S\simeq 1.6$,
respectively. Thus,  we have observed that  parameter-space for
our model II is similar to de Sitter inflation in which $C_0^S=0$,
$C_1^S=0$ and $C_2^S=2$. In relation to the reconstruction of the
background variables, for simplicity we have given  a particular
coupling function $F(\varphi)\propto \varphi^2$, with which we
have found the reconstruction for the scale factor $a(\eta)$ and
the effective potential $V(\varphi)$, respectively (see
Fig.(\ref{fig2})).

Finally in this article, we have not addressed the reconstruction
of the background variables and cosmological perturbations to
another ratios  $z_s/z_t$ in terms of the conformal time, see
Eq.(\ref{gs}). We hope to return to this point in the near future.

\begin{acknowledgments}
J.A.B. was supported by COMISION NACIONAL DE CIENCIAS Y TECNOLOGIA
through FONDECYT Grant N$_0$  11170083. C.G. was supported by
Proyecto DIUDA  REGULAR N$_{0}$  22347. R.H. was supported by
Proyecto VRIEA-PUCV N$_{0}$ 039.309/2018.

\end{acknowledgments}


\end{document}